\documentclass{article}

\usepackage{amsmath}
\usepackage{graphicx}

\begin{document}

\title{Geometric and projection effects in Kramers-Moyal analysis}

\author{Steven J. Lade}

\maketitle

\begin{abstract}
Kramers-Moyal coefficients provide a simple and easily visualized method with which to analyze stochastic time series, particularly nonlinear ones. One mechanism that can affect the estimation of the coefficients is geometric projection effects. For some biologically-inspired examples, these effects are predicted and explored with a non-stochastic projection operator method, and compared with direct numerical simulation of the systems' Langevin equations. General features and characteristics are identified, and the utility of the Kramers-Moyal method discussed. Projections of a system are in general non-Markovian, but here the Kramers-Moyal method remains useful, and in any case the primary examples considered are found to be close to Markovian.
\end{abstract}

\section{Introduction}
In deterministic nonlinear systems, embedding theorems provide one method of reconstructing the system's phase space and thereby obtaining a qualitative understanding, at least, of the system's structure. In stochastic systems no such embedding theorems exist \cite{Kantz_2003}. One approach for analyzing stochastic time series has been pioneered by Friedrich et al. \cite{Friedrich_PRL_1997,Freidrich_PLA_2000} and continues to be developed by Kleinhans \cite{Kleinhans_PLA_2007,Kleinhans_PLA_2005,Kleinhans_PRE_2007} and others. The Kramers-Moyal coefficients (in one dimension),
\begin{equation}
\label{eq:KM}
D^{(n)}(x_0,t) = \lim_{\tau \to 0} \frac{1}{\tau n!} \langle [x(t + \tau) - x_0]^n | x(t)=x_0 \rangle,
\end{equation}
are a critical part of the theory of Fokker-Planck equations, but only recently did Friedrich et al. suggest they could be computed directly from data. The first and second coefficients are known as the drift and diffusion coefficients; the higher-order coefficients are often zero \cite{Risken}. The angle brackets denote an ensemble average, but if the Kramers-Moyal coefficient is constant in time this can be replaced with a time average. If the system can be modeled by a Langevin equation, or, equivalently, a Fokker-Planck equation, then the drift and diffusion coefficients completely characterize of the system.

Such an approach, already applied to fields including neuroscience \cite{Freidrich_PLA_2000}, cardiology \cite{Kuusela_PRE_2004}, traffic engineering \cite{Kriso_PLA_2002}, finance \cite{Friedrich_PRL_2000, Ghasemi_PRE_2007} and turbulence \cite{Friedrich_PRL_1997}, appears well suited to molecular biology, where biological structures commonly move in an overdamped, strongly Brownian environment \cite{Lade_bio_inprep}. Due to structural constraints, however, biological systems are often constrained in their motion to diffuse on a manifold which is no longer Cartesian. Experimentally, meanwhile, position measurements are usually taken in a Cartesian framework. Due to these geometrical effects the Kramers-Moyal coefficients derived from measurements are not the same as on the particle's manifold. We show how these projected drift and diffusion coefficients can be used to make inferences about the original system's dynamics.

For illustration, we consider three specific examples. Motivated by the diffusional search phase of myosin-V \cite{Dunn_NSMB_2007,Shiroguchi_S_2007} and other molecular motors, we consider a particle undergoing tethered diffusion with a rigid tether, so that the particle is constrained to a sphere. We assume its motion is recorded only in one linear dimension, with an experiment like the traveling wave tracking method of Cappello et al. \cite{Cappello_PNAS_2007} This involves the projection of a two-dimensional system onto one dimension. In this example, the diffusion on the sphere is free; in a second example, we make it harmonic in angle from a preferred orientation (`biased diffusion on a sphere'). Lastly, we consider the more complicated example of a particle undergoing free diffusion on a sphere about a point itself undergoing biased diffusion on a sphere. We refer to this as `compound diffusion'.

We begin in section II with a simple example to illustrate the projection concept. After presenting a general framework for predicting geometrical effects in section III, we continue with the less trivial examples in sections IV and V. The effects of a non-Markovian projection are discussed in section VI, followed by conclusions in section VII.

\section{Free diffusion on a sphere}
\label{sec:freediffusion}
Consider a particle diffusing on a sphere, of radius $r$, under potential $U$ and constant local diffusion coefficient $D$. Raible and Engel \cite{Raible_AOC_2004} found that the stochastic differential equations (SDEs, or Langevin equations) describing this motion in spherical polar co-ordinates $(\theta, \phi)$ are
\begin{align}
\label{eq:Langevinthetaphi}
\begin{split}
d\theta &= (-\partial_\theta U + D \cot \theta) dt + \sqrt{2 D} d w_\theta \\
d\phi &= -\frac{\partial_\phi U}{\sin^2\theta} dt + \frac{\sqrt{2 D}}{\sin\theta} dw_\phi.
\end{split}
\end{align}
$dw_\theta$ and $dw_\phi$ are Wiener processes and the SDE is to be interpreted under the It\=o convention. Setting $U = 0$ for free diffusion and projecting onto $z = r\cos\theta$, we have by the It\=o rule for change of variables
\begin{equation}
dz = -2Dz dt - \sqrt{2D(r^2 - z^2)} dw_\theta. \label{eq:freediffusion_proj}
\end{equation}
This corresponds to drift and diffusion coefficients
\begin{align}
\begin{split}
D^{(1)}(z) &= -2Dz \\
D^{(2)}(z) &= D(r^2 - z^2).
\end{split}
\label{eq:projsphere}
\end{align}
Therefore even though in local rectangular co-ordinates on the surface of the sphere the drift is zero and the diffusion constant (`free diffusion'), in this projection the drift is linear, and the diffusion is an inverted (convex) parabola.

The drift coefficient is analogous to a position-dependent force, and the diffusion to a position-dependent temperature or `noise' term \cite{Kriso_PLA_2002}. Although the original motion on the surface was force-free with constant diffusion coefficient, we see the geometric projection introduces an effective `restoring force', proportional to $-z$. It also introduces an effective noise which is maximal at $z = 0$ but decreases to zero at $z = \pm r$: this prevents the projected particle from moving past the edge of the sphere.

The equilibrium probability distribution of this process is \cite{Risken}
\begin{equation*}
%\label{eq:eqmdistr}
P(z) \propto \frac{1}{D^{(2)}(z)} \exp \int^z \frac{D^{(1)}(z)}{D^{(2)}(z)} dz.
\end{equation*}
For free diffusion on a sphere, this gives a uniform distribution
\begin{equation}
\label{eq:freeeqmdistr}
P(z) =
\begin{cases}
\frac{1}{2r}, &|z| < r \\
0, &|z| \ge r
\end{cases}
.
\end{equation}

This illustrates the power of the Kramers-Moyal approach. A na\"ive analysis of a stationary stochastic time series would be to construct its probability distribution. For the example above, this would give a uniform distribution, from which a reasonable guess for the system would be free diffusion in a one dimensional infinite potential well (corresponding to constant diffusion, and drift zero and infinite inside and at the edges of the well, respectively). On computing the drift and diffusion coefficients from a time series by Eq. \eqref{eq:KM}, however, one would recover the drift and diffusion coefficients of Eqs. \eqref{eq:projsphere}, and obtain a different characterization. One explanation of these drift and diffusion coefficients could be to recognize the system as a projection of free diffusion on a sphere, although this identification cannot be unique since we are projecting onto one dimension a two-dimensional system, and without {\it a priori} knowledge could be three- or higher-dimensional. Finally, if the projection is Markovian, as discussed below, the drift and diffusion coefficients fully characterize the dynamics in this co-ordinate, which the probability distribution does not.

Unlike stochastic autoregressive-type models \cite{Kantz_2003}, we see that the Kramers-Moyal approach permits arbitrary nonlinearity in the drift and diffusion coefficients to be recovered, and easily visualized. 

\section{General analysis of projections}
\label{sec:genproj}
Suppose we project the dynamics of a system $\vec{x}(t)$ onto some (collective) variable $X(t) = f(\vec{x}(t))$. Suppose the system can be written as a set of first-order stochastic differential equations. With the use of It\=o's chain rule, an SDE for $X$ in terms of $\vec{x}$ may be derived,
\begin{equation}
\label{eq:stepproject1}
dX = g_X(\vec{x},t) dt + h_X(\vec{x},t) dw.
\end{equation}
We consider the estimation of the Kramers-Moyal coefficients \eqref{eq:KM} for this process. Writing the conditional expectation of Eq. \eqref{eq:KM} in an alternative form gives
\begin{equation*}
D^{(n)}_X(\vec{x}_0,t) = \lim_{\tau \to 0} \frac{1}{n! \tau} \int (X-X_0)^n P(X,t+\tau|\vec{x}_0,t) dX,
\end{equation*}
where $X_0 = f(\vec{x}_0)$. We find $D^{(1)}_X(\vec{x},t) = g_X(\vec{x},t)$, $D^{(2)}_X(\vec{x},t) = \frac{1}{2} h^2_X(\vec{x},t)$, and $D^{(n)}_X(\vec{x},t) = 0$ for $n \ge 3$ \cite{Risken}. But these coefficients still involve the full system variables $\vec{x}$, to which, we assume, the experiment does not have access.

From basic laws of conditional probability, it can be shown that the projection of these Kramers-Moyal coefficients from $\vec{x}$ onto the subspace $X$ (which we distinguish from the unprojected coefficients by its arguments) is
\begin{equation}
D^{(n)}_X(X_0,t) =  \int D^{(n)}_X(\vec{x}_0,t) \frac{\delta(X_0 - f(\vec{x}_0)) P(\vec{x}_0,t)}{P(X_0,t)} d\vec{x}_0. \label{eq:Dproj}
\end{equation}
This constitutes a projection operator for the Kramers-Moyal coefficients,
\begin{equation*}
\mathcal{P}_X \left[ D_X^{(n)}(\vec{x},t) \right] = \int D_X^{(n)}(\vec{x},t) \frac{\delta(X - f(\vec{x})) P(\vec{x},t)}{P(X,t)} d\vec{x}.
\end{equation*}
Clearly $\mathcal{P}_X\mathcal{P}_X = \mathcal{P}_X$ so $\mathcal{P}_X$ is indeed a projection. As $D^{(n)}(\vec{x},t)$ is a dynamical variable on the full phase space $\vec{x}$, the $\mathcal{P}_X$ above coincides with the projection operator from the standard Zwanzig-Mori projection techniques \cite{Zwanzig_2001}. To reiterate, however: before applying the projection operator to an arbitrary Langevin equation, one must transform the equation into the form of \eqref{eq:stepproject1}, one in the variable into which the system is to be projected. The quotient $\delta(X - f(\vec{x})) \frac{P(\vec{x},t)}{P(X,t)}$ is also known as the conditional distribution for the subset of the full phase space distribution $P(\vec{x},t)$ on the surface $f(\vec{x}) = X$. The projection also has implications for the Markov property, as will be discussed in section \ref{sec:Markov}.

If the 1-point densities $P(\vec{x}_0,t)$ and $P(X_0,t)$ are well-defined, so are the projections $D^{(n)}_X(X,t)$ (although they in some cases may be trivially zero). Furthermore, since $D^{(n)}_X(\vec{x},t) = 0$ for $n \ge 3$, then so do $D^{(n)}_X(X,t)$. The projected drift and diffusion coefficients, the equations above show, may be time-dependent even if the original system is homogeneous (the coefficients in its equations of motion independent of time) if the distribution $P(\vec{x}_0,t)$ is time-dependent (non-stationary). If the original system is both homogeneous and stationary, the projected drift and diffusion coefficients will be stationary.
%April 17: Trivially zero for a Kramers equation with 0.1 damping. But add noise to x-variable, PDxx now non-zero but system still non-Markovian. So not trivially zero <=> system is non-Markovian

The Kramers-Moyal coefficients may of course be computed directly from experimental data \eqref{eq:KM} of the projected variable. The necessary averages may be computed by time averaging if the projected drift and diffusion coefficients are stationary, which as just noted is when the full system is homogeneous and stationary. (This is in contrast to Kramers-Moyal analysis on the full system, where the system need only be homogeneous.) This projection operator method, however, provides a means of predicting the projected drift and diffusion coefficients from a model of the full system without the need for direct numerical simulation.

We now illustrate these observations with two examples.

\section{Biased diffusion on a sphere}
\label{sec:biaseddiffusion}
Suppose a particle is again undergoing diffusion on a sphere with local diffusion coefficient $D$, but with a potential harmonic with angle away from a preferred binding direction. Align the $z$-axis with this preferred direction so that $U = k\theta^2/2$. Substituting into the Langevin equations \eqref{eq:Langevinthetaphi} for diffusion on a sphere, the equilibrium solution is
\begin{equation*}
P_{\theta\phi}(\theta) \propto \sin\theta e^{-k\theta^2/2D},
\end{equation*}
normalized such that $\iint P_{\theta\phi}(\theta) d\theta d\phi = 1$.

Let us project this motion onto an axis $s$ tilted at angle $\theta_0$ from the $z$-axis ($\theta = 0$ direction) in the $x-z$ plane,
\begin{align}
s(\theta,\phi) &= z \cos\theta_0 + x \sin\theta_0 \notag \\
&= r\cos\theta_0\cos\theta + \sin\theta_0\sin\theta\cos\phi. \label{eq:s}
\end{align}
The differential
\begin{equation}
ds \equiv g_s(\theta,\phi) dt + h_s(\theta,\phi) dw
\label{eq:biaseddiffusion_proj}
\end{equation}
can be calculated using It\=o's law for change of co-ordinates, and adding the noise terms, which we assume to be independent, in quadrature.

For the steady-state probability distribution as a function of $s$, we calculate
\begin{align*}
P_s(s_0) &= \langle \delta(s_0 - s(\theta,\phi)) \rangle_{\theta,\phi} \\
&= \int \sum_{\phi_0:s(\theta,\phi_0) = s_0} \frac{P_{\theta\phi}(\theta)}{|s_\phi(\theta,\phi_0)|} d\theta,
%\begin{cases}
%2, & \textrm{if } \exists \phi_0 : q(\phi_0) = 0 \\
%0, & \textrm{otherwise}
%\end{cases}
\end{align*}
the second equality by integrating over $\phi$, and where we denote $s_\phi(\theta,\phi_0) \equiv \partial_\phi s(\theta,\phi)|_{\phi=\phi_0}$. The single integral that results we calculate numerically.

Similarly, from Eq. \eqref{eq:Dproj} we project the Kramers-Moyal coefficients by
\begin{multline}
\label{eq:semianalytical}
D_s^{(n)}(s_0) = \frac{1}{P_s(s)} \int \sum_{\phi_0:s(\theta,\phi_0) = s_0} D_s^{(n)}(\theta,\phi_0) \frac{P_{\theta\phi}(\theta)}{|s_\phi(\theta,\phi_0)|} d\theta,
\end{multline}
where we have performed the integral over $\phi$ analytically, and in which $\phi_0$ is itself a function of $\theta$.

\begin{figure}
\includegraphics[width=8.5cm]{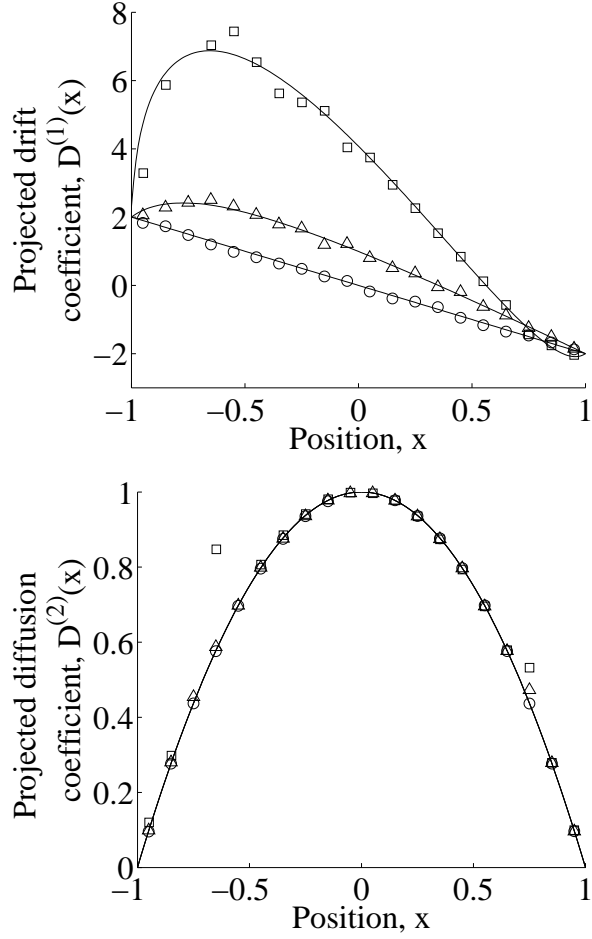}
\caption{Projected (top) drift and (bottom) diffusion coefficients for biased diffusion on a sphere. Semi-analytical predictions \eqref{eq:semianalytical} are shown with solid lines, together with results from direct numerical simulation \eqref{eq:Langevinthetaphi} for restoring force strength $k = 0$ (circles), 1 (triangles) and 5 (squares). Preferred binding angles (and reference for co-ordinate transformation) were $\theta_0 = \pi/2, \pi/4, \pi/4$ respectively. The diffusion coefficient on the surface was $D = 1$ and the radius of the sphere (length of the tether) $r = 1$ in all cases.}
\label{fig:diffonsphereproj}
\end{figure}

In Fig. \ref{fig:diffonsphereproj} we show drift and diffusion coefficients for $s(t)$ computed from direct numerical simulation of Eqs. \eqref{eq:Langevinthetaphi} and by the semi-analytical approach culminating in Eq. \eqref{eq:semianalytical}. The agreement between numerical and theoretical approaches is excellent, with two exceptions. At $x = \pm \cos\theta_0$ there are spikes due to the $\theta_0 = 0, \pi$ singularities in Eqs. \eqref{eq:Langevinthetaphi}. Raible and Engel \cite{Raible_AOC_2004} transform their equations into three-dimensional Cartesian co-ordinates before performing computations; we retain the spherical polar co-ordinates for transparency. Secondly, the larger binding potential for $k = 5$ means the large negative regions of $x$ are poorly sampled and also may lead to finite sampling time effects \cite{Lade_finitetime_inprep}.

From Fig. \ref{fig:diffonsphereproj} we see the potential on the sphere $U = k\theta^2/2$ has an effect additional to the geometric effective force arising from the projection, and it is largest away from the potential minimum $z = \cos\theta_0$. The projected diffusion coefficients, meanwhile, show no change with the strength of the potential on the surface, $k$. We infer that only the geometry of the system has affected the diffusion coefficient, and therefore that an inverted parabola shape for the diffusion coefficient is a good indicator of diffusion on a sphere.

More generally, if a particle is diffusing on a surface that is radially symmetric about the measurement axis $s$, with radius $r(s)$, then the projected diffusion coefficient is independent of the potential on the surface. It can be shown from Raible and Engel's formulas that the projected diffusion coefficient will be
\begin{equation*}
D^{(2)}(s) = \frac{D}{1+(\partial r /\partial s)^2},
\end{equation*}
assuming that the diffusion coefficient on the surface has constant value $D$.

\section{Compound diffusion on a sphere}
We extend the previous scenario to one where there is a second rod undergoing tethered diffusion about the end of the first. We assume the first rod has a preferred binding angle, but that the second rod is diffusing freely. This models the motion of the unbound head during the diffusional search phase of molecular motors such as myosin-V.

In a first approximation, we assume that the motion of the end of the first rod with respect to its tethered end is given by the biased diffusion on a sphere of section IV, and the motion of the free end of the second rod with respect to the end joined to the first rod is given by the free diffusion on a sphere of section II. Once again we project onto a linear co-ordinate. The position of the end of the first rod is then $s(\theta,\phi)$ from Eq. \eqref{eq:s}, and the end of the second rod $X(\theta,\phi,x_2) = s(\theta,\phi) + x_2$, with $x_2$ given by Eq. \eqref{eq:freediffusion_proj} for $z$. The equilibrium distribution of $x_2$, $P_{x2}(x_2)$ is therefore given by Eq. \eqref{eq:freeeqmdistr}. We further assume that the processes $s(\theta,\phi)$ and $x_2$ are independent. Therefore the (unprojected) drift and diffusion coefficients are
\begin{align*}
g_X(\theta,\phi,x_2) &= g_s(\theta,\phi) + g_{x2}(x_2) \\
h_X(\theta,\phi,x_2) &= \sqrt{h^2_s(\theta,\phi) + h^2_{x2}(x_2)}.
\end{align*}

We first require the equilibrium probability distribution for $X$. Integrating $\langle \delta(X - s(\theta,\phi) - x_2) \rangle_{\theta,\phi,x_2}$ over $x_2$ leaves the convolution
\begin{equation*}
P_X(X) = \iint P_{\theta\phi}(\theta) P_{x2}(X-s(\theta,\phi)) d\theta d\phi,
\end{equation*}
which can be computed numerically. For the drift and diffusion coefficients, from Eq. \eqref{eq:Dproj} we obtain, again after integrating over $x_2$,
\begin{multline*}
D^{(n)}_X(X) = \frac{1}{P_X(X)}  \iint D^{(n)}_X(\theta,\phi,X-s(\theta,\phi)) \\
P_{\theta\phi}(\theta,\phi) P_{x2}(X-s(\theta,\phi)) d\theta d\phi.
\end{multline*}

\begin{figure}
\includegraphics[width=8.5cm]{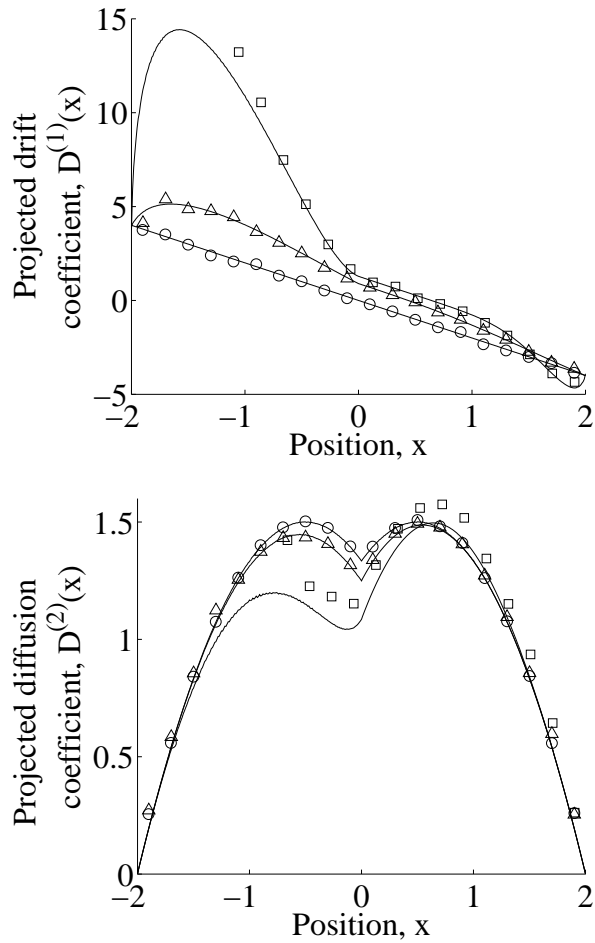}
\caption{Semi-analytical and numerical estimates for projected (top) drift and (bottom) diffusion coefficients for compound diffusion on a sphere, as described in the text. Parameters for the first rod are the same as for Fig. \ref{fig:diffonsphereproj}, and with the same marker shapes. Surface diffusion coefficient and radius for the second rod (with respect to the first) are also $D = 1$ and $r = 1$, respectively.}
\label{fig:compounddiffonsphereproj}
\end{figure}

Numerical results again together with semi-analytical predictions are shown in Fig. \ref{fig:compounddiffonsphereproj}. Observe the unusual bimodal shape for the diffusion coefficient at $k = 0$, reflecting the two `spheres' making up the diffusion process. As $k$ increases this shape begins changing towards a single inverted parabola which one would expect in the limit $k \to \infty$ where the first rod is fixed. The drift is linear at $k = 0$. It maintains an approximately linear region but at larger $k$ develops large swings beyond $x = \cos\theta_0 \pm 1 \approx -0.3, +1.7$, regions which will eventually become forbidden with a fixed first rod ($k \to \infty$).

When the motor is in the diffusional search state, 1-D Kramers-Moyal analysis on experiments tracking the unbound head of myosin-V, or similar molecular motors, should (in the absence of other effects) give results like the above. Alternatively, if the marker is connected to a point on the unbound neck domain, the above simulations should be repeated with a shorter length for the second rod. Yildiz et al. \cite{Cappello_PNAS_2007} have performed such experiments on kinesin, from which they identified the step size and type (hand over hand), but their large sampling interval yields insufficient data for the present analysis.

\section{Markov property}
\label{sec:Markov}
In our examples, the full system is Markov, that is, its next state depends only on the current one and not any other history:
\begin{equation*}
P(\vec{x}_2,t_2|\vec{x}_1,t_1) = P(\vec{x}_2,t_2|\vec{x}_1,t_1;\vec{x}_0,t_0),
\end{equation*}
where $t_2 > t_1 > t_0$. Even if the full system is Markov, however, in general a projection is not \cite{Zwanzig_2001}. Then the projected variable cannot be fully modeled by a Langevin equation (with history-independent coefficients), and so the drift and diffusion coefficients, or even all the Kramers-Moyal coefficients, cannot fully characterize the dynamics of the projected variables. Some authors \cite{Kantz_2003} have criticized the Kramers-Moyal method on this basis, since experimentally one invariably can only access projections, not the entire system.

First, we wish to emphasize that even for projections that are demonstrably not Markovian, the Kramers-Moyal coefficients remain well-defined. In this case one may not write down an equation of motion of the system from the reconstructed coefficients, but they remain a useful tool for characterization and comparison of time series. Conversely, the existence of Kramers-Moyal coefficients, or 2-point probability densities, of course do not by themselves infer anything about the Markov property of the underlying process.

\begin{figure}
\includegraphics[width=8.5cm]{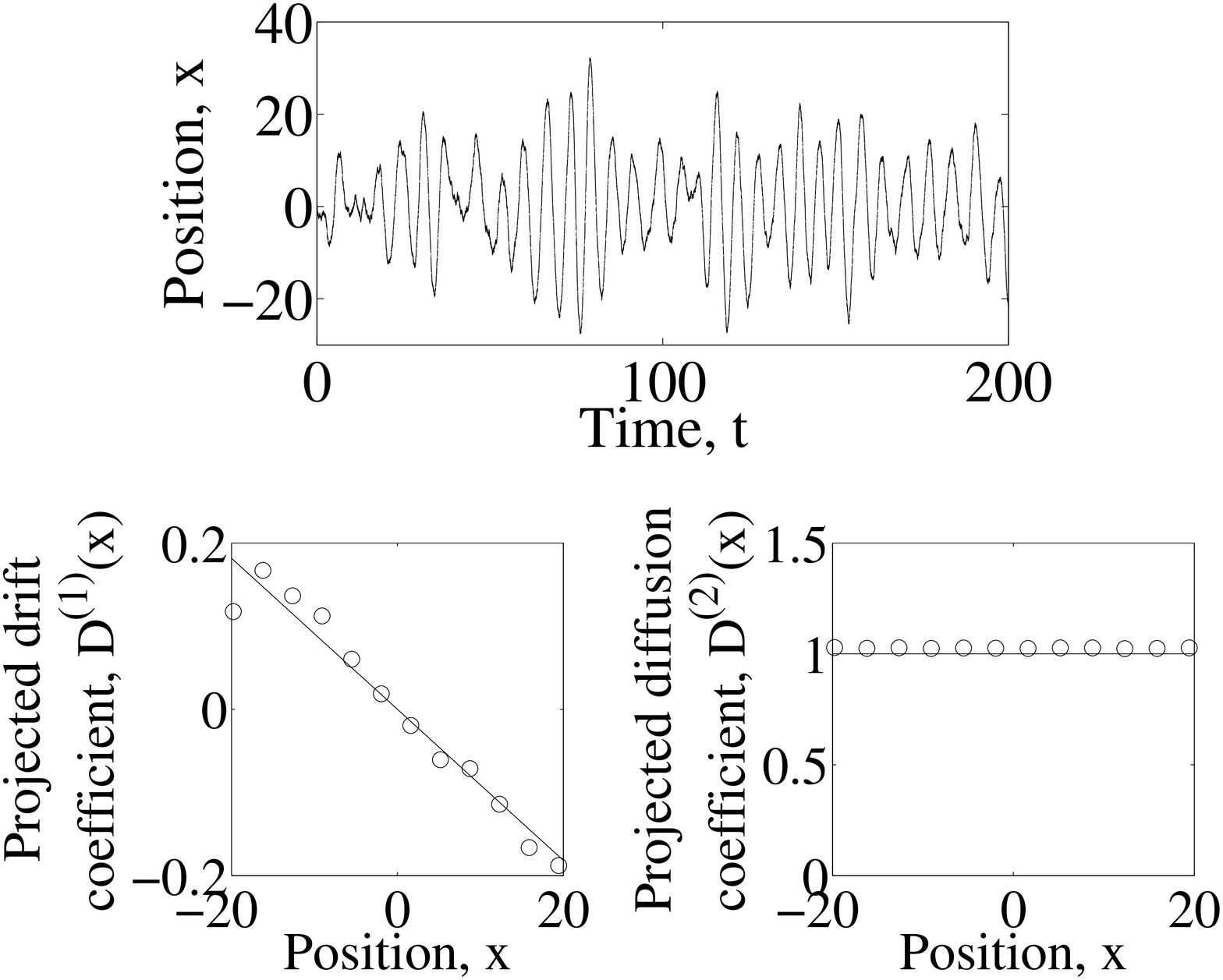}
\caption{Sample trajectory (top) and projected (left) drift and (right) diffusion coefficients for the projection $X = x_1$ of the system of Eqs. \eqref{eq:OU2D} with $a = 0$, $b = 1$, $c = -1$, $d = -0.1$, $D_1 = 1$ and $D_2 = 10$ by direct numerical simulation (circles) and analytically (lines). Analytical predictions were calculated using Erickson's results \cite{Erickson_AMS_1971} for multi-dimensional Ornstein-Uhlenbeck processes, giving $D^{(1)}(X) = \frac{(a + d)(ad-bc) D_1}{(ad-bc+d^2)D_1 + b^2D_2} X$ and $D^{(2)}(X) = D_1$, under suitable stability conditions on the SDEs \eqref{eq:OU2D}.}
\label{fig:OU2D}
\end{figure}

As an example of the Kramers-Moyal coefficients in a non-Markovian projection, consider the two-dimensional Ornstein-Uhlenbeck process
\begin{align}
\label{eq:OU2D}
\begin{split}
dx_1 = (a x_1 + b x_2) dt + \sqrt{2D_1} dW_1 \\
dx_2 = (c x_1 + d x_2) dt + \sqrt{2D_2} dW_2
\end{split}
\end{align}
projected onto $X = x_1$. With a specific set of parameters, we plot the drift and diffusion coefficients in Fig. \ref{fig:OU2D}. As we will show below, this projection is non-Markovian. For this system the projection operation reduces to $\int D^{(n)}_X(X,x_2) P(X,x_2) dx_2 / P(X)$. Qualitatively, the projected drift and diffusion coefficients for a value of the projected variable $X = x_1$ are the  values of the unprojected coefficients averaged by the likely locations of $x_2$ given this $x_1$. In other words, they are the expectations over the conditional probability density $P(x_2|x_1)$. For this system we see in Fig. \ref{fig:OU2D} an effective deterministic force tending to restore $X$ towards the origin, that is confine it to a finite region, and that there is a noise source operating directly on the $X$ variable.

%As stated by Eq. \label{eq:Dproj}, in the non-Markovian case the projected Kramers-Moyal coefficients for a given value of the projected variable will be an average of the full system's Kramers-Moyal coefficients, weighted by the probability (conditional on the value of the projected variable) of the full system being in a particular state. 

There are some cases where we know the projection will be Markovian. If the projected variable is {\it slow} compared to the other degrees of freedom of the system, the projection will be Markovian \cite{Zwanzig_2001}. In the unbiased diffusion on a sphere of section \ref{sec:freediffusion} we saw that because of the special symmetry of the system, the second of the two degrees of freedom ($\phi$) was irrelevant for the projected variable, which by Eq. \eqref{eq:freediffusion_proj} we see is Markovian.

\begin{figure}
\includegraphics[width=8.5cm]{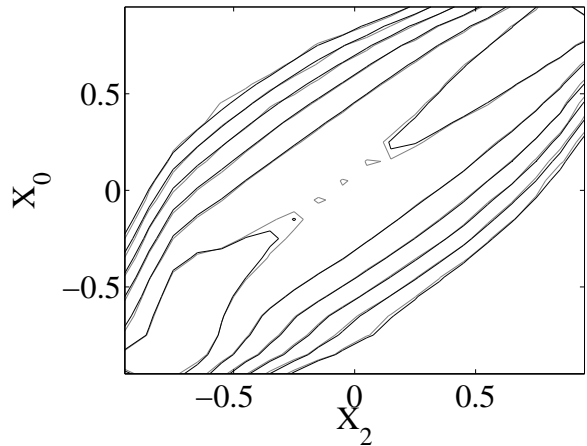}
\caption{Chapman-Kolmogorov equation for projected biased diffusion on a sphere \eqref{eq:biaseddiffusion_proj} with $\tau = 0.1$. Contours, in logarithmic scale, of the left-hand side of the equation are in gray and right-hand side in black.}
\label{fig:Markov_CK}
\end{figure}

\begin{figure}
\includegraphics[width=8.5cm]{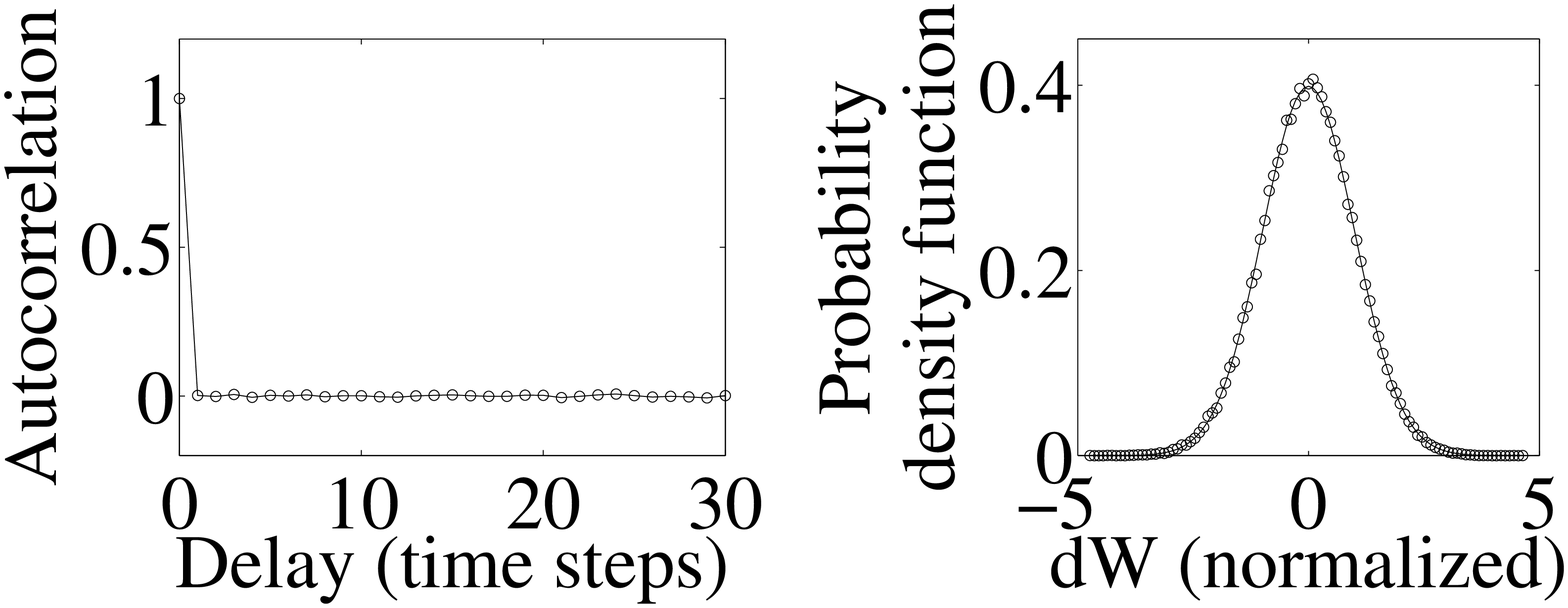}
\caption{(left) Normalized autocorrelation and (right) probability density (circles) with Gaussian fit (solid line) of the reconstructed noise, as per the the {\it a posteriori} test of Micheletti et al. \cite{Micheletti_JCP_2008}, for projected biased diffusion on a sphere \eqref{eq:biaseddiffusion_proj}. The test reconstructs the effective noise $dW$ from the reconstructed drift and diffusion coefficients, assuming a Langevin equation of the form $dX = g(X)dt + h(X)dW$ holds. If $dW$ has delta-autocorrelation and Gaussian probability density, claims the test, the process is Markov.}
\label{fig:Markov_Micheletti}
\end{figure}

The Chapman-Kolmogorov equation \cite{Risken}
\begin{equation*}
P(X_2,t_2|X_0,t_0) = \int P(X_2,t_2|X_1,t_1) P(X_1,t_1|X_0,t_0) dX_1,
%\label{eq:CK}
\end{equation*}
although not sufficient for the Markov property, is generally taken as a good test of it \cite{Friedrich_PRL_2000, Ghasemi_PRE_2007, Ragwitz_PRL_2001}. We tested the biased diffusion on a sphere of section \ref{sec:biaseddiffusion} with the Chapman-Kolmogorov equation and the {\it a posteriori} test of Micheletti et al. \cite{Micheletti_JCP_2008}. Stationarity was assumed and the transition probabilities were calculated by time average, setting $t_2 - t_1 = t_1 - t_0 = \tau$. The results are shown graphically in Figs. \ref{fig:Markov_CK} and \ref{fig:Markov_Micheletti}. These tests indicate that this projection of biased diffusion on a sphere is at least approximately Markovian. To a good approximation, then, the projected variable can be modeled by a Langevin equation in that variable, and in which the noise is Gaussian and delta-correlated.

\begin{figure}
\includegraphics[width=8.5cm]{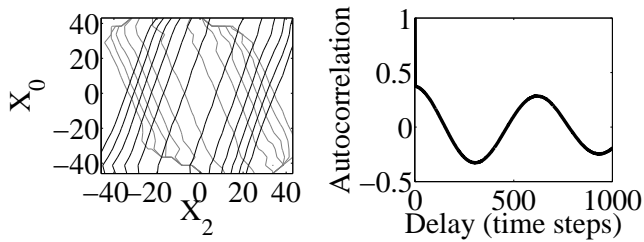}
\caption{(left) Chapman-Kolmogorov equation with $\tau = 1$ for the projection onto $X = x_1$ of the Ornstein-Uhlenbeck process \eqref{eq:OU2D} with parameters as in Fig. \ref{fig:OU2D}. Contours, in logarithmic scale, of the left-hand side are in gray and right-hand side in black. They clearly do not coincide. (right) Autocorrelation of the reconstructed noise, as per the Micheletti test described in Fig. \ref{fig:Markov_Micheletti}. The noise is clearly not delta-autocorrelated.}
\label{fig:nonMarkov}
\end{figure}

For comparison we show in Fig. \ref{fig:nonMarkov} clear failure of the Chapman-Kolmogorov equation and Micheletti tests for a non-Markovian system, that of Eq. \eqref{eq:OU2D} projected onto $X = x_1$.

\section{Conclusions}
We began with a simple example of projecting free diffusion on a sphere onto a rectangular co-ordinate. A general projector operator formalism for projecting Kramers-Moyal coefficients was then established. This was applied to biologically-inspired examples of biased diffusion on a sphere, and compound diffusion on a sphere, where excellent agreement with numerical results was observed. Generally speaking, the diffusion coefficient is more useful for identifying for the geometry of the surface on which the object is diffusing, the the drift coefficient for the potential on the surface. It was noted that the Kramers-Moyal coefficients can provide useful information even when the projection is not Markovian, which in general is the case, although the examples considered here were close to Markovian.

\bibliography{myosin}

\begin{thebibliography}{10}

\bibitem{Kantz_2003}
Holger Kantz and Thomas Schreiber.
\newblock {\em Nonlinear Time Series Analysis}.
\newblock Cambridge University Press, 2003.

\bibitem{Friedrich_PRL_1997}
R.~Friedrich and J.~Peinke.
\newblock Description of a turbulent cascade by a {Fokker-Planck} equation.
\newblock {\em Phys. Rev. Lett.}, 78(5):863--866, Feb 1997.

\bibitem{Freidrich_PLA_2000}
R.~Friedrich, S.~Siegert, J.~Peinke, St. L\"uck, M.~Siefert, M.~Lindemann,
  J.~Raethjen, G.~Deuschl, and G.~Pfister.
\newblock Extracting model equations from experimental data.
\newblock {\em Phys. Lett. A.}, 271:217--222, 2000.

\bibitem{Kleinhans_PLA_2007}
D.~Kleinhans and R.~Friedrich.
\newblock Maximum likelihood estimation of drift and diffusion functions.
\newblock {\em Phys. Lett. A}, 368:194--198, 2007.

\bibitem{Kleinhans_PLA_2005}
D.~Kleinhans, R.~Friedrich, A.~Nawroth, and J.~Peinke.
\newblock An iterative procedure for the estimation of drift and diffusion
  coefficients of {Langevin} processes.
\newblock {\em Phys. Lett. A}, 346(1-3):42--46, 2005.

\bibitem{Kleinhans_PRE_2007}
David Kleinhans, Rudolf Friedrich, Matthias W\"{a}chter, and Joachim Peinke.
\newblock Markov properties in presence of measurement noise.
\newblock {\em Phys. Rev. E}, 76(4):041109, 2007.

\bibitem{Risken}
H.~Risken.
\newblock {\em The Fokker Planck Equation: Methods of Solution and
  Applications}.
\newblock Springer-Verlag, Berlin, 1984.

\bibitem{Kuusela_PRE_2004}
Tom Kuusela.
\newblock Stochastic heart-rate model can reveal pathologic cardiac dynamics.
\newblock {\em Phys. Rev. E}, 69(3):031916, Mar 2004.

\bibitem{Kriso_PLA_2002}
S.~Kriso, J.~Peinke, R.~Friedrich, and P.~Wagner.
\newblock Reconstruction of dynamical equations for traffic flow.
\newblock {\em Phys. Lett. A}, 299(2-3):287--291, 2002.

\bibitem{Friedrich_PRL_2000}
R.~Friedrich, J.~Peinke, and Ch. Renner.
\newblock How to quantify deterministic and random influences on the statistics
  of the foreign exchange market.
\newblock {\em Phys. Rev. Lett.}, 84(22):5224--5227, May 2000.

\bibitem{Ghasemi_PRE_2007}
Fatemeh Ghasemi, Muhammad Sahimi, J.~Peinke, R.~Friedrich, G.~Reza Jafari, and
  M.~Reza~Rahimi Tabar.
\newblock Markov analysis and {Kramers-Moyal} expansion of nonstationary
  stochastic processes with application to the fluctuations in the oil price.
\newblock {\em Phys. Rev. E}, 75(6):060102(R), 2007.

\bibitem{Lade_bio_inprep}
S.~J. Lade, E.~M. Craig, and H.~Linke.
\newblock Kramers-{M}oyal analysis of myosin-{V} walking data.
\newblock in progress.

\bibitem{Dunn_NSMB_2007}
A.~R. Dunn and J.~A. Spudich.
\newblock Dynamics of the unbound head during myosin {V} processive
  translocation.
\newblock {\em Nature Struct. Mol. Biol.}, 14:246--248, 2007.

\bibitem{Shiroguchi_S_2007}
Katsuyuki Shiroguchi and Kazuhiko {Kinosita Jr.}
\newblock Myosin {V} walks by lever action and {B}rownian motion.
\newblock {\em Science}, 316(5828):1208--1212, 2007.

\bibitem{Cappello_PNAS_2007}
G.~Cappello, P.~Pierobon, C.~Symonds, L.~Busoni, J.~C.~M. Gebhardt, M.~Rief,
  and J.~Prost.
\newblock {Myosin V} stepping mechanism.
\newblock {\em Proc. Nat. Acad. Sci. USA}, 104:15328--15333, 2007.

\bibitem{Raible_AOC_2004}
M.~Raible and A.~Engel.
\newblock Langevin equation for the rotation of a magnetic particle.
\newblock {\em Appl. Organometal. Chem.}, 18:536--541, 2004.

\bibitem{Zwanzig_2001}
R.~Zwanzig.
\newblock {\em Nonequilibrium Statistical Mechanics}.
\newblock Oxford University Press, New York, 2001.

\bibitem{Lade_finitetime_inprep}
S.~J. Lade.
\newblock Finite sampling time effects in kramers-moyal analysis.
\newblock submitted.

\bibitem{Erickson_AMS_1971}
R.~V. Erickson.
\newblock Constant coefficient linear differential equations driven by white
  noise.
\newblock {\em Ann. Math. Stat.}, 42:820, 1971.

\bibitem{Ragwitz_PRL_2001}
Mario Ragwitz and Holger Kantz.
\newblock Indispensable finite time corrections for {Fokker-Planck} equations
  from time series data.
\newblock {\em Phys. Rev. Lett.}, 87(25):254501, Dec 2001.

\bibitem{Micheletti_JCP_2008}
C.~Micheletti, G.~Bussi, and A.~Laio.
\newblock Optimal {Langevin} modeling of out-of-equilibrium molecular dynamics
  simulations.
\newblock {\em J. Chem. Phys.}, 129:074105, 2008.

\end{thebibliography}
\bibliographystyle{unsrt}
\end{document}